\documentclass[12pt]{article}
\pdfoutput=1
\usepackage{graphicx}
\usepackage{amsmath}
\usepackage{amssymb}

\begin{document}
 
\def\eqrefc{\eqref}
\def\wt{\widetilde}
\def\a{\alpha}
\def\b{\beta}
\def\c{\varepsilon}
\def\d{\delta}
\def\e{\epsilon}
\def\f{\phi}
\def\g{\gamma}
\def\h{\theta}
\def\k{\kappa}
\def\l{\lambda}
\def\m{\mu}
\def\n{\nu}
\def\p{\psi}
\def\q{\partial}
\def\r{\rho}
\def\s{\sigma}
\def\t{\tau}
\def\u{\upsilon}
\def\v{\varphi}
\def\w{\omega}
\def\x{\xi}
\def\y{\eta}
\def\z{\zeta}
\def\D{\Delta}
\def\G{\Gamma}
\def\H{\Theta}
\def\L{\Lambda}
\def\F{\Phi}
\def\P{\Psi}
\def\S{\Sigma}

\def\o{\over}
\def\beq{\begin{eqnarray}}
\def\eeq{\end{eqnarray}}
\newcommand{\gsim}{ \mathop{}_{\textstyle \sim}^{\textstyle >} }
\newcommand{\lsim}{ \mathop{}_{\textstyle \sim}^{\textstyle <} }
\newcommand{\vev}[1]{ \left\langle {#1} \right\rangle }
\newcommand{\bra}[1]{ \langle {#1} | }
\newcommand{\ket}[1]{ | {#1} \rangle }
\newcommand{\EV}{ {\rm eV} }
\newcommand{\KEV}{ {\rm keV} }
\newcommand{\MEV}{ {\rm MeV} }
\newcommand{\GEV}{ {\rm GeV} }
\newcommand{\TEV}{ {\rm TeV} }
\def\diag{\mathop{\rm diag}\nolimits}
\def\Spin{\mathop{\rm Spin}}
\def\SO{\mathop{\rm SO}}
\def\O{\mathop{\rm O}}
\def\SU{\mathop{\rm SU}}
\def\U{\mathop{\rm U}}
\def\Sp{\mathop{\rm Sp}}
\def\SL{\mathop{\rm SL}}
\def\tr{\mathop{\rm tr}}

\def\IJMP{Int.~J.~Mod.~Phys. }
\def\MPL{Mod.~Phys.~Lett. }
\def\NP{Nucl.~Phys. }
\def\PL{Phys.~Lett. }
\def\PR{Phys.~Rev. }
\def\PRL{Phys.~Rev.~Lett. }
\def\PTP{Prog.~Theor.~Phys. }
\def\ZP{Z.~Phys. }


\baselineskip 0.7cm

\begin{titlepage}

\begin{flushright}
IPMU11-0207\\
ICRR-Report-601-2011-18
\end{flushright}

\vskip 1.35cm
\begin{center}
{\large \bf The Lightest  Higgs Boson Mass\\ in\\ Pure Gravity Mediation Model
}
\vskip 1.2cm
Masahiro Ibe$^{1,2}$ and Tsutomu T. Yanagida$^{1}$
\vskip 0.4cm
$^1${\it ICRR, University of Tokyo, Kashiwa 277-8582, Japan}\\
$^2${\it IPMU, TODIAS, University of Tokyo, Kashiwa 277-8583, Japan}

\vskip 1.5cm

\abstract{
We discuss the lightest Higgs boson mass in the minimal 
supersymmetric Standard Model with ``pure gravity mediation".
By requiring that the model provides the observed dark matter density,
we find that the lightest Higgs boson  is predicted to be below 
132\,GeV. 
We also find that the upper limit on the lightest Higgs boson mass becomes
 $128$\,GeV, if we further assume thermal leptogenesis mechanism 
as the origin of baryon asymmetry of universe.
The interrelations between the Higgs boson mass and the 
gaugino masses are also discussed.
}
\end{center}
\end{titlepage}

\setcounter{page}{2}

\section{Introduction}
Supersymmetry is the most attractive candidate for beyond the Standard Model.
Surprisingly, the assumption of 
spontaneous breaking of supersymmetry  (SUSY)
is enough to give rise to the masses of the superpartners of the
Standard Model particles in the framework of supergravity. 
Scalar bosons acquire  SUSY breaking soft masses 
at the tree level\,\cite{Nilles:1983ge} and gauge fermions (gauginos) 
at the one-loop level\,\cite{Giudice:1998xp, Randall:1998uk, hep-ph/9205227}.
We call this minimal setup as ``pure gravity mediation". 
The most attractive feature of this framework is that 
we do not need any additional fields for
the mediation of SUSY breaking effects.

If we assume that the pure gravity mediation model is within the reach of 
the LHC experiments, the scale of spontaneous SUSY breaking is chosen 
to be around $10^{11-12}$\,GeV so that the gaugino masses generated 
at the one-loop level are in the hundreds GeV to the TeV range.
Interestingly, the purely gravity mediated model with this mass range
has many attractive features compared to the conventional models 
owing to the minimal setup. 
First of all, there is no serious Polonyi problem\,\cite{Polonyi},%
\footnote{See also Ref.\,\cite{hep-ph/0605252} for 
the Polonyi problem in dynamical supersymmetry breaking models.} 
since there is no Polonyi field required to generate the gaugino masses.
The cosmological gravitino problem\,\cite{kkm} is also solved  in this setup.
This is because the gravitino mass is in the hundreds TeV range 
and decays before the Big-Bang Nucleosynthesis (BBN).
The problems of flavor-changing neutral currents and CP violation in the supersymmetric Standard Model 
become very mild thanks to relatively large masses for squarks and sleptons. 
Furthermore, we have a good candidate of dark matter 
in the universe\,\cite{Gherghetta:1999sw,hep-ph/9906527, Ibe:2004tg,ArkaniHamed:2006mb}.
Especially, it was pointed out in Ref.\,\cite{Ibe:2004tg} that the pure gravity mediation model 
has a wide rage of parameter space consistent with the thermal leptogenesis\,\cite{leptogenesis}. 
The unification of the gauge coupling constants at the very high energy scale 
also provides a strong motivation to the model.

Encouraged by these advantages, 
we discuss the mass of the lightest Higgs boson
in the minimal SUSY Standard Model (MSSM).
We find the upper limits on the lightest 
Higgs boson mass is predicted to be about $132$\,GeV.
The requirement of the successful leptogenesis lowers
the upper limit down to about $128$\,GeV.
These predictions will be tested soon at the LHC experiments. 

The organization of the paper is as follows.
In section\,\ref{sec:PGM} and \ref{sec:Higgs}, 
we discuss the masses of  the MSSM superparticles and the lightest Higgs boson
in the pure gravity mediation model.
In section\,\ref{sec:HiggsBound}, we derive the upper limits on the lightest Higgs boson 
mass by requiring the consistent dark matter density.
We also discuss the consistency of the model with thermal leptogenesis.
In section\,\ref{sec:Gaugino}, we discuss 
the interrelation between the lightest Higgs boson mass and the gaugino masses.
The final section is devoted to our conclusions.

\section{Purely Gravity Mediated SUSY Breaking}\label{sec:PGM}
\subsection*{Sfermions and Gauginos}
In the pure gravity mediation model,
the only new ingredient other than the MSSM fields
is  a (dynamical) SUSY breaking sector. 
Then, the soft SUSY breaking masses of squarks, sleptons and Higgs bosons are mediated
by the supergravity effects at the tree-level. 
With a generic K\"ahler potential, all the scalar bosons obtain the SUSY breaking masses of
the order of the gravitino mass, $m_{3/2}$.  
For the gaugino masses, on the other hand, tree-level contributions 
in the supergravity are extremely suppressed 
since we have no  SUSY breaking fields which are singlet
under any symmetries.

At the one-loop level, however, the gaugino masses are generated by the supergravity 
effects without having singlet SUSY breaking 
fields\,\cite{Giudice:1998xp,Randall:1998uk,hep-ph/9205227}.  
The one-loop generated so-called anomaly-mediated gaugino masses are given by
\begin{eqnarray}
    M_{a}  = -\frac{b_{a}g_{a}^{2}}{16\pi^{2}} m_{3/2}\ ,
    \label{eq:GauginoMass}
\end{eqnarray}
where $a$ denotes the three standard-model gauge groups ($a=1,2,3$),
$g_{a}$ gauge coupling constants, and $b_{a}$ coefficients of
the renormalization-group equations of $g_{a}$, i.e. $b_{a}= (-33/5,
-1, 3)$.
Therefore, the framework of the pure gravity mediation does
not require any new mediator fields to make the superparticles massive.

The important feature  of the anomaly-mediated gaugino spectrum
is that the lightest gaugino is the neutral wino.
The charged wino is slightly heavier  
than the neutral one by about $155$\,MeV$-170$\,MeV
due to one-loop gauge boson contributions\,\cite{Feng:1999fu}. 
Thus, it is quite tempting to explore whether 
the neutral wino can be a candidate for dark matter.
In fact, thermal relic density of the wino is consistent with
the observed dark matter density for $M_2 \simeq 2.7$\,TeV\,\cite{hep-ph/0610249,arXiv:0706.4071}.
The relatively large mass of thermal wino dark matter 
stems from the large annihilation cross section of the winos into $W$-bosons.
The lighter wino than $2.7$\,TeV is also a good candidate
once the relic abundance is provided by the non-thermal production
by the late time decay of the gravitinos which were produced when
the universe had high 
temperature\,\cite{Gherghetta:1999sw,hep-ph/9906527, Ibe:2004tg,ArkaniHamed:2006mb}.
As we will discuss, the consistent mass range of the wino dark matter puts 
upper limit on the lightest Higgs boson mass in the pure gravity mediation model.

\subsection*{Higgs Sector}
In the purely gravity mediated models,
we also expect that the two additional mass parameters in the Higgs sector,
the so-called $\mu$- and $B$-parameters, 
are also of the order of the gravitino mass.
Indeed, without any special symmetries, 
we expect the following K\"ahler potential,
\begin{eqnarray}
    K \ni c H_{u}H_{d} + 
    \frac{c'}{M_{PL}^{2}} Z^{\dagger} Z H_{u}H_{d} + h.c..
\end{eqnarray}
Here, $Z$ is a chiral superfield in the hidden sector, which may
or may not be a composite field, $M_{PL}$ is the reduced Planck scale, 
and $c$ and $c'$ are coefficients of ${ O}(1)$.%
\footnote{Even if $Z$ is a composite field, $c'$ can be 
${\cal O}(1)$.}
The above K\"ahler potential leads to the $\mu$- and the
$B$-parameters\,\cite{Inoue:1991rk}
\begin{eqnarray}
    \label{eq:Muterm}
    \mu_H &=& c m_{3/2},\\
    \label{eq:Bterm}
    B \mu_H &=& c m_{3/2}^{2} + c'\frac{|F_X|^2}{M_{PL}^{2}},
\end{eqnarray}
where $F_Z$ is the vacuum expectation value of the $F$-component of
$Z$.%
\footnote
{We assume that the vacuum expectation value of $Z$ is much smaller
than $M_{PL}$.}
Thus, $\mu$- and $B$-parameters are both expected to be of ${ O}(m_{3/2})$,
and hence, the higgsinos are expected to be as heavy as  the sfermions and the gravitino.

For successful electroweak symmetry breaking, 
one linear combination of the Higgs bosons should be light
which is denoted by $ h = \sin\b H_{u} - \cos\b H_{d}^{*}$
with a mixing angle $\beta$.
Here, $H_u$ and $H_d$ are up- and down-type Higgs bosons, respectively.  
In terms of the mass parameters, the mixing angle is given by 
\begin{eqnarray}
\label{eq:angle}
  \sin2\beta = \frac{2 B\mu_H}{m_{H_u}^2+ m_{H_d}^2 + 2|\mu_H|^2} \ ,  
\end{eqnarray}
while the light Higgs boson requires a tuning between mass parameters,
\begin{eqnarray}
   (|\mu_H|^2+m_{H_u}^2)    (|\mu_H|^2+m_{H_u}^2) - (B\mu_H)^2 \simeq 0\ ,
\end{eqnarray}
at the energy scale of the heavy scalars.
Therefore, by remembering that squared masses of $H_u$ and $H_d$,
$m_{H_{u,d}}^2$, as well as $B$ and $\mu_H$ are of the order of the gravitino mass,
the mixing angle $\beta$ 
is expected to be of ${ O}(1)$.%
\footnote{Hereafter, we treat the $\mu_H$ and $B$ parameters 
as real valued parameters just for simplicity, although 
our discussions are not changed even if they are complex valued.
}

In summary of the pure gravity mediation, 
the mass spectrum and the Higgs mixing angle are expected to be;
\begin{itemize}
\item The sfermions and the gravitino are in the ${\cal O}(10^{4-6})$\,GeV range.
\item The higgsinos and the heavier Higgs bosons are in the ${\cal O}(10^{4-6})$\,GeV range.
\item The gauginos are in the hundreds to thousands TeV range.
\item The Higgs mixing angle is of order of unity, i.e. $\tan\beta = {\cal O}(1)$.
\end{itemize}

Notice that the pure gravity mediation model has some similarities to
the Split Supersymmetry\,\cite{hep-th/0405159,hep-ph/0406088,hep-ph/0409232} 
for $M_{\rm SUSY} \simeq 10^{4-6}$\,GeV.
The important difference is that we do not expect $M_{\rm SUSY} \gg 10^{4-6}$\,GeV,
since we rely on the anomaly-mediated gaugino masses 
in the pure gravity mediation model.%
\footnote{See discussions on the possible cancellation of the 
anomaly-mediated gaugino masses\,\cite{hep-ph/0409232,Izawa:2010ym}.
}
In this sense, the pure gravity mediation model is more close to
the PeV-scale Supersymmetry\,\cite{hep-ph/0411041}
and the Spread Supersymmetry\,\cite{arXiv:1111.4519}.
The other important and more practical difference is the size of $\mu$-term.
In the Split Supersymmetry, it is assumed that the higgsinos are also in the TeV range,
while we consider they are as heavy as the gravitino.
Therefore, we can distinguish our scenario from the Split Supersymmetry 
by searching for the higgsinos at the collider experiments.

\section{The Lightest Higgs Boson Mass}\label{sec:Higgs}
Below the scale of the heavy scalars, $M_{\rm SUSY} = {\cal O}(m_{3/2})$,
the Higgs sector consists of the light Higgs boson $h$ whose potential is given
by,
\begin{eqnarray}
 V(h) = \frac{\lambda}{2}(h^\dagger h-v^2)^2\ ,
\end{eqnarray}
where $v \simeq 174.1$\,GeV is determined to reproduce the observed 
$Z$ boson mass.
At the tree-level, the Higgs coupling constant $\lambda$ satisfies 
the so-called the SUSY relation,
\begin{eqnarray}
\label{eq:SUSY}
  \lambda = \frac{1}{4} \left(\frac{3}{5}g_1^2+ g_2^2 \right) \cos^22\beta\ .
\end{eqnarray}
This is the famous and remarkable feature of the MSSM where 
the physical Higgs boson mass, $m_h^2 = 2 \lambda v^2$,
is not a free parameter but a prediction of the model.

Below $M_{\rm SUSY}$, the above SUSY relation is violated 
by the SUSY breaking effects through the radiative corrections\,\cite{TU-363}.
The first contribution to  deviates the SUSY relation
is the radiative correction through the renormalization-group equation.
At the one-loop level, the renormalization-group equation is roughly given by
\begin{eqnarray}
 \frac{d \lambda}{d t} \sim \frac{12}{16\pi^2} ( \lambda^2  + \lambda y_t^2- y_t^4) \ ,
\end{eqnarray}
where $y_t$ denotes the top Yukawa coupling, and we have neglected
gaugino couplings for illustrative purpose. 
By imposing the SUSY relation in Eq.\,(\ref{eq:SUSY}) at the renormalization scale $Q=M_{\rm SUSY}$,
the renormalization-group equation can be approximately solved by,
\begin{eqnarray}
 \lambda(m_h) \sim \lambda(M_{\rm SUSY})  + \frac{12}{(4\pi)^2}y_t^4 \ln \frac{M_{\rm SUSY}}{m_h}\ .
\end{eqnarray}
Therefore, we expect that the physical Higgs mass receives a large positive 
correction for $M_{\rm SUSY} ={\cal O}(10^{4-6})$\,GeV.

The second contribution which deviates the SUSY relation comes 
from the finite correction to the Higgs quartic coupling from the trilinear 
couplings. 
At the one-loop level, this contribution is given by,
\begin{eqnarray}
\label{eq:mhmax}
\delta \lambda &\simeq & \frac{6}{(4\pi)^2} y_t^4 
\left( \frac{X_t^2}{m_{\tilde t}^2}-
\frac{1}{12}\frac{X_t^4}{m_{\tilde t}^4}\right) , \cr
X_t &=& {A_t - \mu_H \cot\b}\, \simeq -\mu_H \cot\beta \ , \cr
m_{\tilde t}^2 &=& m_{t_L}^2 + m_{t_R}^2  \ ,
\end{eqnarray}
where $A_t$ is the trilinear coupling constant between Higgs and stops,
and $m_{t_{L,R}}^2$ denote the squared soft masses of the left and right stops.
Notice that  $A_t$ is expected to be suppressed at the tree-level of the supergravity.%
\footnote{Here, we again assume that $\vev{Z} \ll M_{\rm PL}$. }
Since $\mu_H$ is in the gravitino mass range and $\tan\beta = O(1)$, 
this correction can be sizable in the pure gauge mediation model.

With these discussions in mind, we 
compute the lightest Higgs boson mass for  given $M_{\rm susy}$,
$\mu_H$ and $\tan\beta$.
In our analysis, we numerically solve the full one-loop renormalization-group
equations of the Higgs quartic coupling, the gauge couplings, the gaugino 
couplings,  the Yukawa couplings of the third generation fermions,
and the gaugino masses  given in Ref.\,\cite{hep-ph/0406088}.
We also include the weak scale threshold corrections to those 
parameters in accordance with Ref.\,\cite{arXiv:0705.1496,arXiv:1108.6077}.
Notice that we decouple the higgsino contributions to the renormalization
group equations at $Q=\mu_H$ and match the coupling
constants below and above that scale,
since $\mu_H$ is much heavier than the TeV scale.

\begin{figure}[t]
\begin{center}
\begin{minipage}{.49\linewidth}
  \includegraphics[width=.9\linewidth]{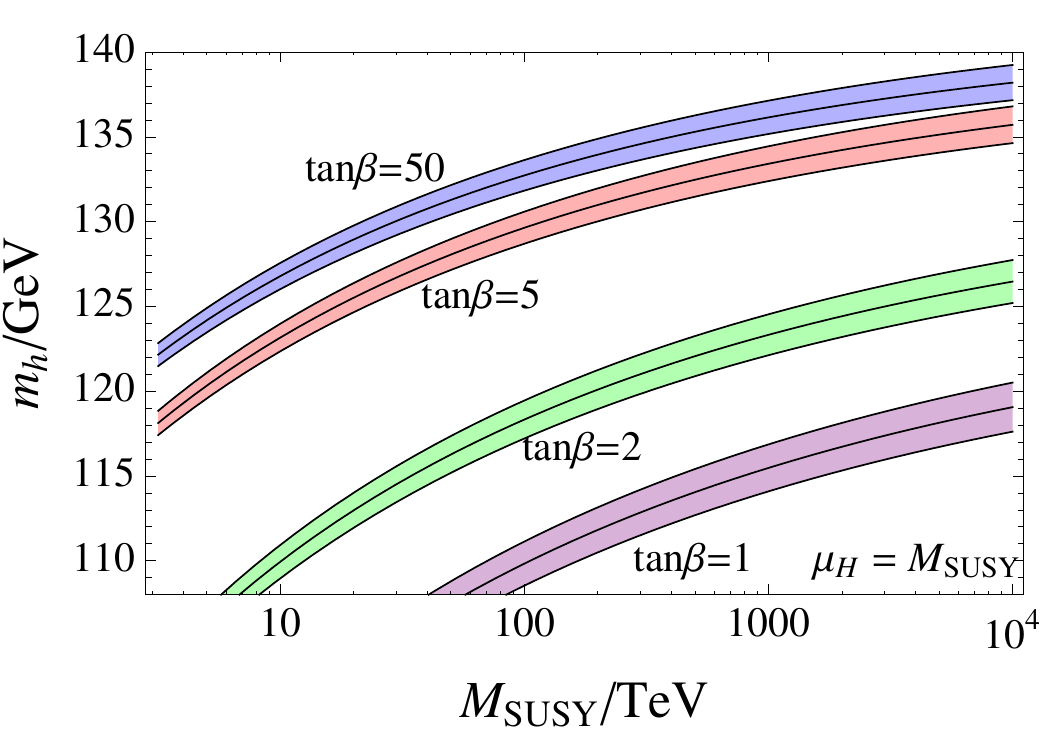}
  \end{minipage}
  \begin{minipage}{.49\linewidth}
  \includegraphics[width=.9\linewidth]{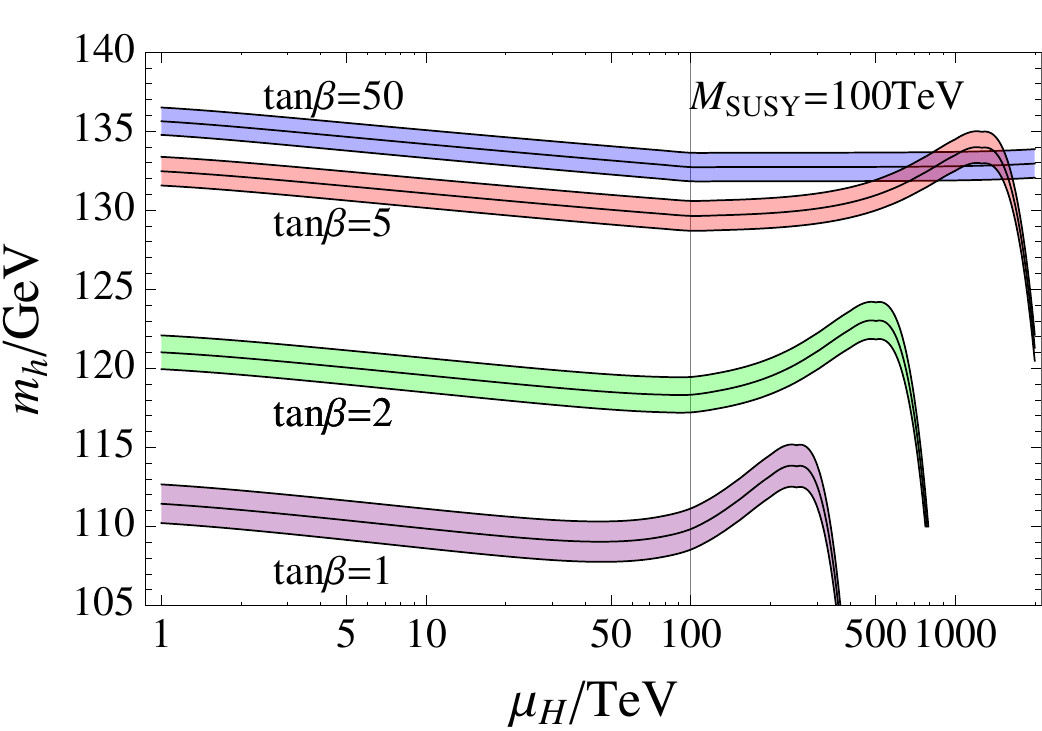}
  \end{minipage}
\caption{\sl \small
Left) The lightest Higgs boson mass as a function of $M_{\rm SUSY}$
with $\mu_H = M_{\rm SUSY}$.
The result is slightly lighter than the one in Ref.\,\cite{arXiv:1108.6077} 
due to the large $\mu$-term (see the right panel).
Right) The lightest Higgs boson mass as a function of $\mu_H$ for $M_{\rm SUSY} = 100$\,TeV.
In both panels, 
the color bands show the $1\sigma$ error of the top quark mass, $m_{\rm top}= 173.2\pm0.9$\,GeV\,\cite{arXiv:1107.5255},
while we have taken the central value of the strong coupling constant,
$\alpha(M_Z)=0.1184\pm 0.0007$\,\cite{arXiv:0908.1135}.
We have also fixed the gaugino masses to
$M_1 = 900$\,GeV, 
$M_2 = 300$\,GeV and $M_3 =- 2500$\,GeV as reference values,
although the predicted Higgs boson mass is insensitive to the gaugino masses.
}
\label{fig:Higgs1}
\end{center}
\end{figure}

In Fig.\,\ref{fig:Higgs1}, we show the parameter dependancies of the  lightest Higgs boson mass.
The left panel of the figure shows the Higgs boson mass 
as a function of $M_{\rm SUSY}$.
In the figure, we have taken $\mu_H = M_{\rm SUSY}$.
The color bands represent 
the $1\sigma$ error on the top quark mass, $m_{\rm top}= 173.2\pm0.9$\,GeV.%
\footnote{We have not shown the uncertainty due 
to the $1\sigma$ error on the strong coupling constant which is smaller
than the one from the top mass error.}
The figure shows that the lightest Higgs boson mass can 
easily exceed the lower bound from the LEP experiments, $m_{h}>114.4$\,GeV\,\cite{hep-ex/0306033}
for $\tan\b=O(1)$.
The lightest Higgs boson mass larger than $120$\,GeV is also easily 
realized for the wide range of parameters.

The right panel shows the $\mu_H$ dependence of the lightest Higgs boson mass
for $M_{\rm SUSY}=  100$\,TeV.
The color bands again correspond to the $1\sigma$ error of the top quark mass.
The figure shows that the lightest Higgs boson mass decreases monotonically for 
the larger $\mu_H$ for relatively small $\mu_H$ region, i.e.  $\mu_H \ll M_{\rm SUSY}$.
This is due to the fact that the gaugino coupling contributions 
increase the Higgs quartic coupling constant at the low energy
via the renormalization group equations.
For $\mu_H = {\cal O}(M_{\rm SUSY})$, on the other hand,
the finite threshold correction to the Higgs quartic coupling in Eq.\,(\ref{eq:mhmax})
becomes important especially for the small $\tan \beta$.
The peaks of the lightest Higgs boson mass correspond to the 
parameters which satisfy $X_t \simeq \sqrt{6} m_{\tilde t}$.

\begin{figure}[t]
\begin{center}
  \includegraphics[width=.4\linewidth]{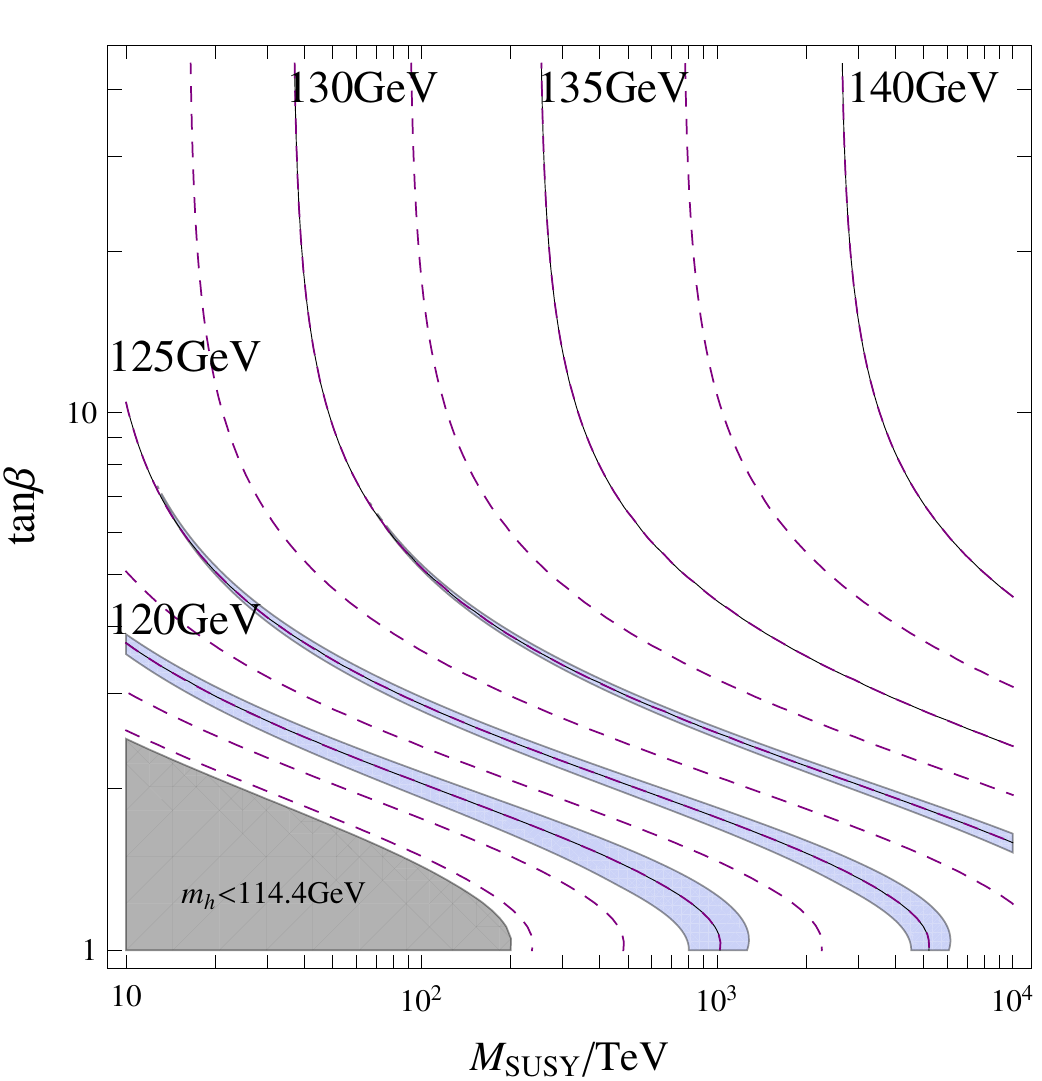}
\caption{\sl \small
The contour plot of the lightest Higgs boson mass.
The bands for $m_h = 120,125,130,135,140$\,GeV
represent the effects of the theoretical uncertainty 
of the ratio $\mu_H/M_{\rm SUSY}$ to the lightest 
Higgs boson mass. 
We have assumed that $M_{\rm SUSY}/3<\mu_H<3M_{\rm SUSY}$.
We have used the central values of the $1\sigma$ errors
of the strong coupling constant and the top quark mass.
}
\label{fig:Higgs2}
\end{center}
\end{figure}

In Fig.\,\ref{fig:Higgs2}, we show the contour plot of the lightest Higgs boson mass as a function of
$M_{\rm SUSY}$ and $\tan\beta$.
In the figure, we have used the central values of the $1\sigma$ errors
of the strong coupling constant and the top quark masses.
For  given parameters, we have used the gaugino masses
which are obtained by solving the full one-loop renormalization group equations
with the anomaly-mediated boundary condition in Eq.(\ref{eq:GauginoMass}) at $Q = M_{\rm SUSY}$
with $m_{3/2} = M_{\rm SUSY}$.
The color bands represent the effects of the theoretical uncertainty 
of the ratio $\mu_H/M_{\rm SUSY}$ on the lightest 
Higgs boson mass. 
We have taken $M_{\rm SUSY}/3 < \mu_H < 3M_{\rm SUSY}$.
The figure shows that the effect of the theoretical uncertainty is sizable
for a small $\tan\beta$ region where  the finite correction in Eq.\,(\ref{eq:mhmax}) to 
the Higgs quartic coupling can be large.

\section{Upper Bound on The Lightest Higgs Boson Mass}\label{sec:HiggsBound}

\begin{figure}[t]
\begin{center}
  \includegraphics[width=.5\linewidth]{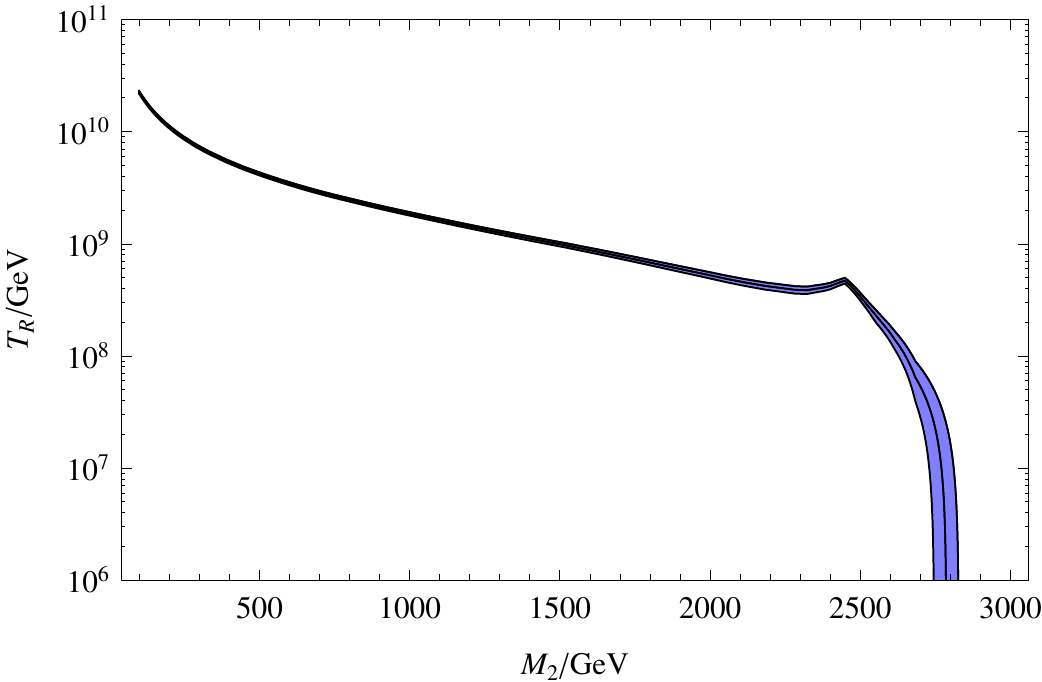}
\caption{\sl \small
The required reheating temperature of universe as a function
of the wino mass for the consistent dark matter density. 
We have used the thermal relic density given in Refs.\,\cite{hep-ph/0610249,arXiv:0706.4071}.
The color bands correspond to the $1\sigma$ error of
the observed dark matter density, $\Omega h^2 = 0.1126\pm 0.0036$\,\cite{arXiv:1001.4538}.
For a detailed discussion see also Ref.\,\cite{Ibe:2004tg}.
}
\label{fig:TR}
\end{center}
\end{figure}
As we mentioned above, the lightest superparticle in the pure gravity mediation 
is the neutral wino which can be a good dark matter candidate.
The important feature of the wino dark matter scenario 
is that the current abundance consists of two contributions.
The one is from the thermal relic density of the wino itself, 
and the other from the the late time decay of the gravitino.
Notice that the late time decay of the gravitino does not 
cause the gravitino problems since the gravitino decay before the 
BBN\,\cite{kkm}. 

The thermal relic density of the wino is determined by the annihilation cross section of
the winos into the $W$-bosons via the weak interaction. 
The resultant  relic density  $\Omega^{(TH)} h^2(M_2)$
can be found in Ref.\,\cite{hep-ph/0610249,arXiv:0706.4071}.
The thermal relic density saturates the observed dark matter density
$\Omega h^2\simeq 0.11$ for $M_2 \simeq 2.7$\,TeV,
while it is quickly decreasing for the lighter wino.
The non-thermal relic density is, on the other hand, proportional to 
the gravitino number density
which is proportional to the reheating 
temperature $T_R$ after inflation,
\begin{eqnarray}
 \Omega^{(NT)} h^2(M_2,T_R) \simeq 0.16 \times \left(  \frac{M_2}{300\,{\rm GeV}} \right)
 \left(  \frac{T_R}{10^{10}\,{\rm GeV}} \right)\ .
\end{eqnarray}
The total relic density is given by,
\begin{eqnarray}
\Omega h^2 = \Omega^{(TH)}(M_2)+  \Omega^{(NT)} h^2(M_2,T_R)\ . 
\end{eqnarray}
Therefore, the wino which is lighter than $2.7$\,TeV can be the dominant component of 
the dark matter for an appropriate reheating temperature.

Fig.\,\ref{fig:TR} shows the required reheating temperature of universe as a function
of the wino mass for the consistent dark matter density. 
The color bands correspond to the $1\sigma$ error of
the observed dark matter density, $\Omega h^2 = 0.1126\pm 0.0036$\,\cite{arXiv:1001.4538}.
It is remarkable that the required reheating temperature is consistent 
with the lower bound on $T_R$ for the successful thermal leptogenesis, $T_R \gtrsim 10^{9.5}$\,GeV\,\cite{leptogenesis}.

\begin{figure}[t]
\begin{center}
\begin{minipage}{.49\linewidth}
  \includegraphics[width=.9\linewidth]{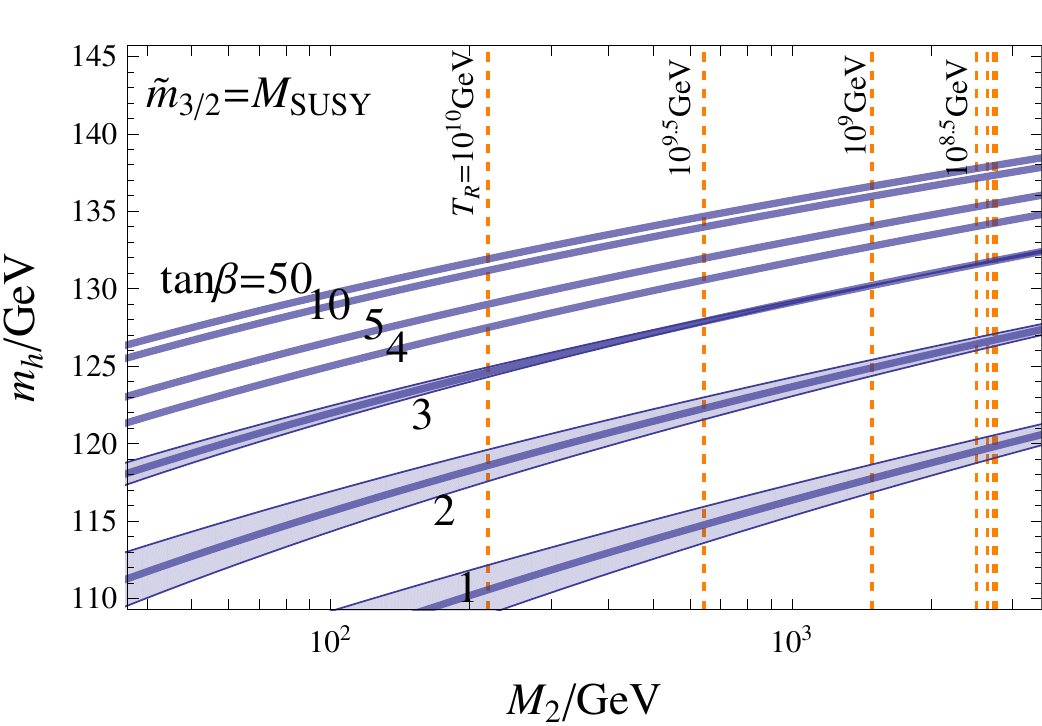}
  \end{minipage}
  \begin{minipage}{.49\linewidth}
  \includegraphics[width=.9\linewidth]{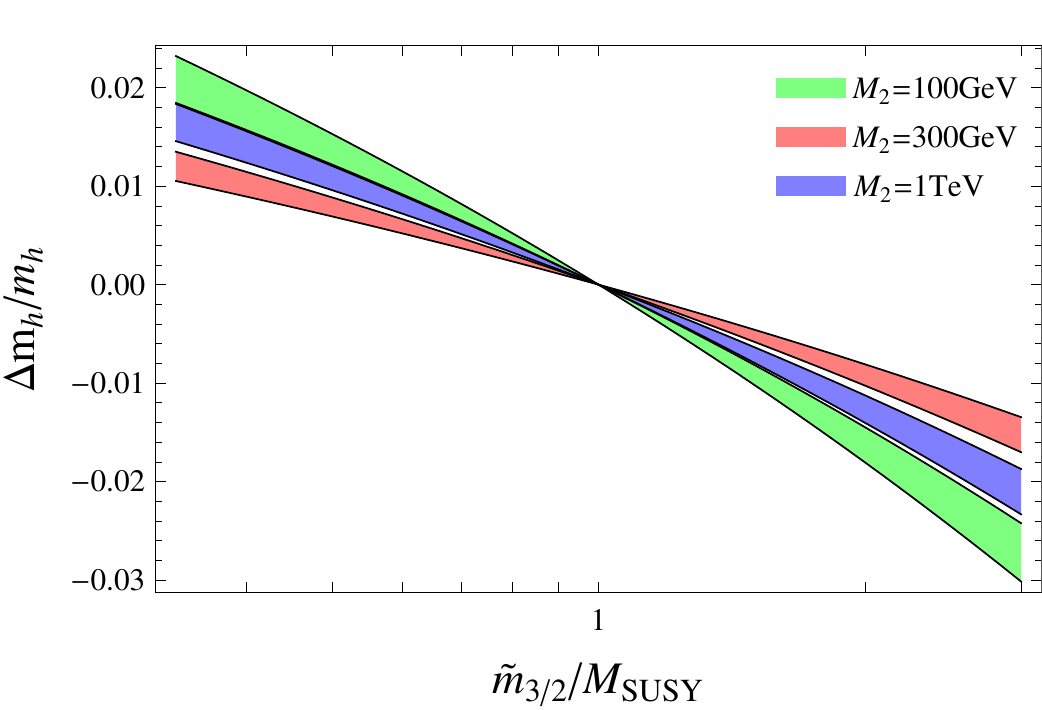}
  \end{minipage}
\caption{\sl \small
Left) The lightest Higgs boson mass for a given wino mass.
We also show the required reheating temperature for the
successful wino dark matter scenario as dashed lines (see Fig.\,\ref{fig:TR}).
Right) The lightest Higgs boson mass dependence on the 
theoretical uncertainty from the ratio $\tilde{m}_{3/2}/M_{\rm SUSY}$.
}
\label{fig:Higgswino}
\end{center}
\end{figure}

Now, let us interrelate the wino dark matter density and the lightest Higgs boson mass.
As we have discussed, the lightest Higgs boson mass is determined 
for given $M_{\rm SUSY} = {\cal O}(m_{3/2})$ and  $\tan \beta$. 
The wino mass is, on the other hand, is given by,
\begin{eqnarray}
\label{eq:numM2}
 M_2 \simeq 3\times 10^{-3}\,m_{3/2}\ ,
\end{eqnarray}
with the anomaly-mediated boundary condition in Eq.\,(\ref{eq:GauginoMass}) 
at $Q = M_{\rm SUSY}$.%
\footnote{
The current experimental bound on $M_2$ is $M_2 \geq 88$\,GeV
obtained at the LEP experiments\,\cite{hep-ex/0203020}.
The mass of the wino dark matter is also constrained
to $M_2 \gtrsim 200-250$\,GeV 
 by the
observed light element abundance through 
the dark matter annihilation at the BBN era\,\cite{arXiv:0810.1892}.
}
Thus, with the theoretical uncertainty of the ratio $m_{3/2}/M_{\rm SUSY}$,
we can interrelate the Higgs boson mass and the wino mass.

In Fig.\,\ref{fig:Higgswino}, we show the lightest Higgs boson mass
as a function of the wino mass for $\tilde m_{3/2} = M_{\rm SUSY}$.
(Here, we have used $\tilde m_{3/2} \simeq m_{3/2}$ instead of $m_{3/2}$.
The definition of $\tilde{m}_{3/2}$ is given in Eq.\,(\ref{eq:effgravitino}).)
The color bands of the left panel again the effects of the theoretical uncertainty 
of the ratio $\mu_H/M_{\rm SUSY}$  as discussed in the previous section.
In the figure, we also show the contour plot of the required 
reheating temperature for the wino dark matter scenario.
The figure shows that the Higgs boson mass is predicted 
to be lighter for the higher reheating temperature for a given $\tan\beta$.

The right panel of the figure shows the dependence 
of the lightest Higgs boson mass on the theoretical uncertainty of the 
ratio, $\tilde{m}_{3/2}/M_{\rm SUSY}$.
The each color band corresponds to $3<\tan\b<50$ for a given vale of  $M_2$.
The smaller $\tan\beta$ is, the larger the effect of the uncertainty is.
The figure shows that the effect of the theoretical uncertainty from the ratio 
$\tilde{m}_{3/2}/M_{\rm SUSY}$ is less  than about $2$\,\% for the wide range of parameters.

\begin{figure}[t]
\begin{center}
  \includegraphics[width=.5\linewidth]{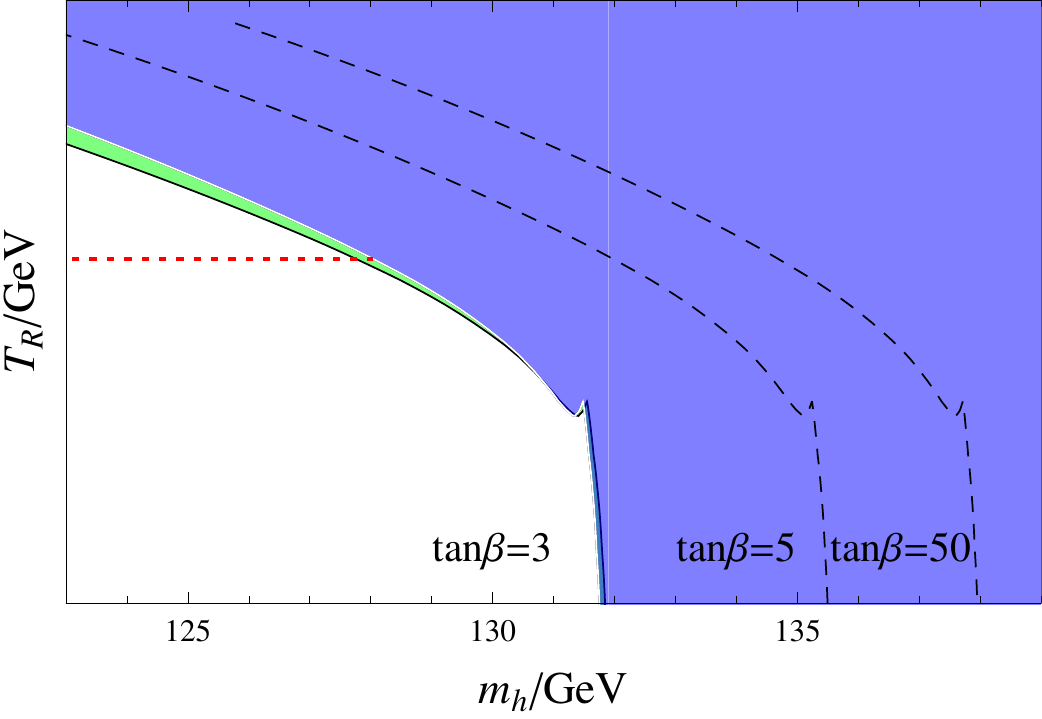}
\caption{\sl \small
The upper limit on the reheating temperature as a function of the
lightest Higgs boson mass.
The green band represents the effects of the theoretically uncertain ratio 
$\mu_H/M_{\rm SUSY}$ which we have taken between $M_{\rm SUSY}/3 < \mu_H < 3 M_{\rm SUSY}$.
The effect of the theoretical uncertainty from the ratio $\tilde{m}_{3/2}/M_{\rm SUSY}$ 
can be read off from the right panel of Fig.\,\ref{fig:Higgswino}.
}
\label{fig:HiggsTR}
\end{center}
\end{figure}

From the Fig.\,\ref{fig:Higgswino}, we can derive the upper limit 
on the reheating temperature after inflation for a given lightest Higgs boson mass.
In  Fig.\,\ref{fig:HiggsTR}, we show the upper limit on $T_R$ for $\tan\beta =3$ which is 
the typical value expected in the pure gravity mediation.
The thin green band represents the effects of the theoretical uncertainty 
from the $\mu_H/M_{\rm SUSY}$ where we have again 
taken $M_{\rm SUSY}/3 < \mu_H < 3 M_{\rm SUSY}$.
We also show the upper limit on the results for $\tan\beta = 5$ and $\tan\beta = 50$
for comparison, although $\tan\beta=50$ is quite unlikely in the pure gravity mediation.

The figure shows that the dark matter constraint puts the upper limit on the Higgs boson mass
is about $m_{h}\simeq 132$\,GeV.
Furthermore, the requirement of thermal leptogenesis puts more stringent constraint on the 
Higgs boson mass down to $m_{h}= 128$\,GeV.
These upper limits will be tested at the LHC experiments very soon.
The effects of the theoretical uncertainties and the $1\sigma$ error
on the top quark masse which are not included this figure can be read off from the 
previous figures.

Before closing this section, let us comment on the threshold corrections
to the gaugino masses at the higgsino threshold
\cite{Giudice:1998xp,Gherghetta:1999sw},
\begin{eqnarray}
    \Delta M_{1}^{(higgsino)} &=& 
    \frac{3}{5} \frac{g_{1}^{2}}{16 \pi^{2}} L,
    \label{eq:DelM1} \\
    \Delta M_{2}^{(higgsino)} &=& 
    \frac{g_{2}^{2}}{16 \pi^{2}} L,
    \\
    \Delta M_{3}^{(higgsino)} &=&  0,
    \label{eq:DelM3}
\end{eqnarray}
where
\begin{eqnarray}
    L \equiv \mu_H \sin2\beta 
    \frac{m_{A}^{2}}{|\mu_H|^{2}-m_{A}^{2}} \ln \frac{|\mu_H|^{2}}{m_{A}^{2}} 
        \label{L-parameter}\ .
\end{eqnarray}
Here, $m_A$ is the mass of heavy Higgs bosons which is given by 
\begin{eqnarray}
m_A^2 = m_{H_u}^2 + m_{H_d}^2 + 2 |\mu_H|^2\ .
\end{eqnarray}
For $\mu_H={\cal O}(m_{3/2})$, $\Delta M_{a}^{(Higgs)}$ (for $a=1,2$)
can be comparable to the anomaly-mediated gaugino masses.
In the above analysis,  we have  introduced an effective gravitino mass scale,
\begin{eqnarray}
\label{eq:effgravitino}
\tilde {m}_{3/2}= m_{3/2} + L ,
\end{eqnarray}
so that $M_2$ is expressed by,
 \begin{eqnarray}
 M_2 = \frac{g_2^2}{16\pi^2} (m_{3/2} + L) = \frac{g_2^2}{16\pi^2} \tilde{m}_{3/2}\ .
\end{eqnarray}
The numerical value of the wino mass for a given $\tilde{m}_{3/2}$
is obtained by replacing $m_{3/2}$ to $\tilde{m}_{3/2}$ in Eq.\,(\ref{eq:numM2}).
Since either $m_{3/2}$ or $\tilde{m}_{3/2}$ 
is expected to be in the same order of $M_{\rm SUSY}$,%
\footnote{If there is a cancellation between $m_{3/2}$ and $L$,
the effective gravitino mass $\tilde{m}_{3/2}$ can be very small 
compared with $M_{\rm SUSY}$, which leads to 
a very large lightest Higgs boson mass for a given wino mass.
We do not consider such cancellation in this paper.
}
we estimated the effects of the 
theoretical uncertainties by sweeping $M_{\rm SUSY}/3 < \tilde{m}_{3/2} < 3 M_{\rm SUSY} $.

\section{Gaugino mass and Higgs boson mass}\label{sec:Gaugino}
Finally, let us briefly discuss the interrelation between the lightest Higgs boson
mass and the gaugino masses.
In the pure gravity mediation, the gauginos are the only superparticles
which can be discovered at the LHC experiments, since
the sfermions  are expected to be as heavy as ${\cal O}(10^{4-6})$\,GeV.%
\footnote{See for example Ref.\,\cite{hep-ph/0610277,arXiv:0705.3086,arXiv:0802.3725,Alves:2011ug}
for the search of the gauginos at the LHC experiments.}
Even worse, the gluino pole mass obtained by the 
anomaly-mediated boundary condition at $Q = M_{\rm SUSY}$
is about $7-10$ times larger than the wino mass.
For example, the gluino mass is about $4$\,TeV 
for $M_2 = 500$\,GeV.
This feature implies that the search of the superparticles at 
the LHC experiments is  very difficult in most parameter space
of the pure gravity mediation.

One possible way out from this pessimistic prediction
can be obtained from the higgsino contributions to the gaugino masses 
in Eqs.\,(\ref{eq:DelM1})-(\ref{eq:DelM3}).
That is, for a given value of $M_2$, the gluino mass is now given by,
\begin{eqnarray}
 M_{3} &\simeq & -(7-10) \times\frac{ M_2}{1+\delta_{\tilde H}}\, ,\cr
 \delta_{\tilde H} &=& 
 \sin2\beta\frac{\mu_H}{m_{3/2}} \frac{m_A^2}{|\mu_H|^2 - m_A^2} \ln\frac{|\mu_H|^2}{m_A^2}
 = {\cal O}(1)\times \sin^22\beta \ .
\end{eqnarray}
In the final expression of $\delta_{\tilde H}$, we have used Eq.\,(\ref{eq:angle}).
Therefore, the gluino mass can be significantly smaller than 
the above mentioned value for  $\tan\beta=O(1)$.%
\footnote{Depending on the sign (or the complex phase) of $\delta_{\tilde H}$,
the gluino can be significantly heavier than the prediction with
the anomaly-mediation boundary condition.}

\begin{figure}[t]
\begin{center}
  \includegraphics[width=.5\linewidth]{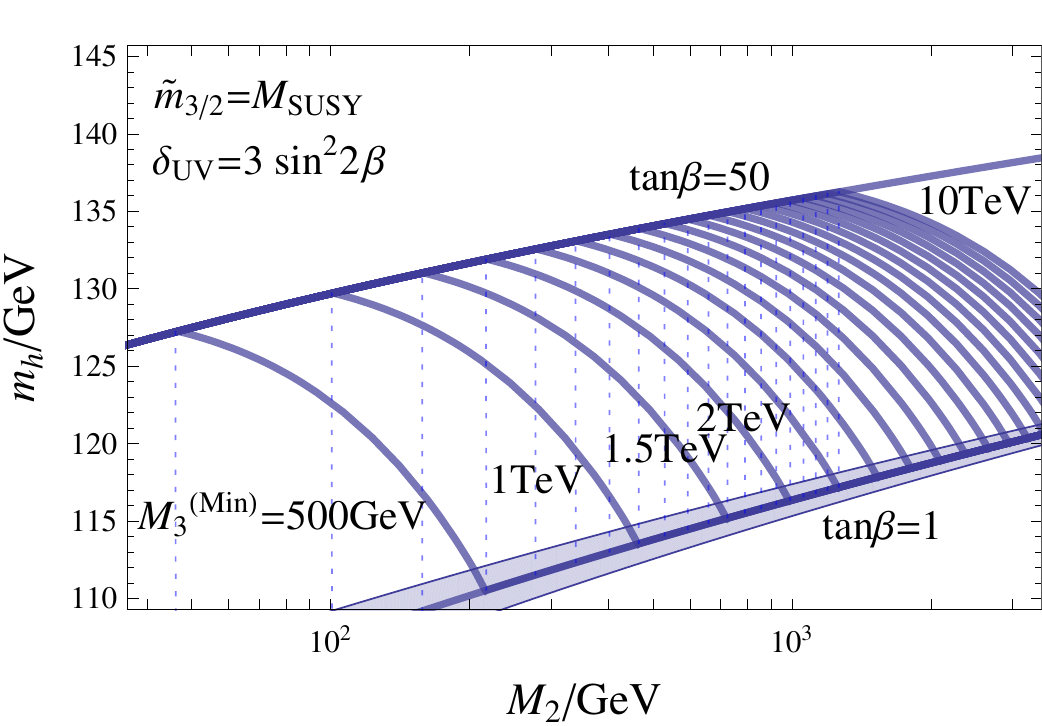}
\caption{\sl \small
The contours
of the lightest possible gluino mass 
as a function of the wino and the  lightest Higgs boson masses.
We have assumed that $\delta_{\tilde H} = 3\sin^22\beta$.
The dashed contours show the gluino mass prediction 
without the higgsino threshold effects.
The effects of the theoretical uncertainties from the ratios $\mu_{H}/M_{\rm SUSY}$
and $\tilde{m}_{3/2}/M_{\rm SUSY}$ can be read from the previous figures.
}
\label{fig:gluino}
\end{center}
\end{figure}

In Fig.\,\ref{fig:gluino}, we show the contour plot of the lightest possible 
gluino mass for  given wino and Higgs boson masses with the higgsino threshold effects
on the wino mass.
Here, we are assuming $\delta_{\tilde H} = 3\sin^22\beta$.
The dotted contours show the gluino mass with the anomaly-mediated boundary conditions
($\delta_{\tilde H}=0$) for comparison. 
The dotted contours are insensitive to the Higgs boson mass.
The figure shows that the gluino can be significantly lighter 
the prediction with the anomaly-mediated boundary condition
for a small $\tan\beta$,
while the effect is vanishing for $\tan\beta ={\cal O}(10)$. 

It should be also noted that we can put 
the lower limit on the lightest possible gluino mass for a given wino mass
once the Higgs mass is determined experimentally.
For example, the figure shows that  the gluino can be as light as $1.5$\,TeV
for $m_{h} \simeq 125$\,GeV and $M_2 \simeq 400$\,GeV.
These features of the pure gravity mediation enhance the testability 
of the model at the LHC experiments.

\section{Conclusions}
In this paper, we discussed 
the lightest Higgs boson mass in the pure gravity mediation model
which consistently provides the observed dark matter density.
The important features of the pure gravity mediation model are
(i) the sfermions, the higgsinos and the gravitinos are as heavy as 
$10^{4-6}$\,GeV (ii) the gaugino masses are in the TeV range
and deviating from the so called GUT relation
(iii) $\tan\beta = {\cal O}(1)$.
With these features, we found the upper limit on the lightest 
Higgs boson mass is predicted to be about $132$\,GeV.
The requirement of the successful leptogenesis lowers
the upper limit down to about $128$\,GeV.
These predictions will be tested at the LHC experiments very soon.

We also discussed the interrelation between the lightest Higgs boson mass
and the gaugino masses.
We found that the gluino mass for given wino and Higgs boson masses 
can be significantly smaller than the predictions with the anomaly-mediated boundary conditions 
due to the higgsino threshold effects on the wino mass.
Therefore, the pure gravity mediation model can be extensively tested by
the interplay between the Higgs searches and the gaugino searches  at the LHC experiments.

In our discussion, we have not studied the constraints on the
wino dark matter scenario from the cosmic ray experiments.
Since the wino has a rather large annihilation cross section
into $W$-boson, it is promising that the model
can be tested through the cosmic ray observations.
The detailed analysis is in preparation.%
\footnote{See for example Ref.\cite{Fujii:2002kr}, for earlier works.}

\section*{Acknowledgements}
We would like to thank S.\,Matsumoto for useful discussions on the wino dark matter property. 
This work was supported by the World Premier International Research Center Initiative
(WPI Initiative), MEXT, Japan.
The work of T.T.Y. was supported by JSPS Grand-in-Aid for Scientific Research (A)
(22244021).


\begin{thebibliography}{99}
\bibitem{Nilles:1983ge}
For  a review, H.~P.~Nilles,
  Phys.\ Rept.\  {\bf 110} (1984) 1.
  
  
\bibitem{Giudice:1998xp}
  G.~F.~Giudice, M.~A.~Luty, H.~Murayama and R.~Rattazzi,
  %
  JHEP {\bf 9812}, 027 (1998).
\bibitem{Randall:1998uk}
  L.~Randall and R.~Sundrum,
  %
  Nucl.\ Phys.\ B {\bf 557}, 79 (1999).
\bibitem{hep-ph/9205227} 
  M.~Dine and D.~MacIntire,
  Phys.\ Rev.\ D\ {\bf 46}, 2594  (1992)
  [hep-ph/9205227].


\bibitem{Polonyi}
  G.~D.~Coughlan, W.~Fischler, E.~W.~Kolb, S.~Raby and G.~G.~Ross,
  Phys.\ Lett.\ B {\bf 131}, 59 (1983).

\bibitem{hep-ph/0605252} 
  M.~Ibe, Y.~Shinbara and T.~T.~Yanagida,
  Phys.\ Lett.\ B\ {\bf 639}, 534  (2006)
  [hep-ph/0605252].
  
  
\bibitem{kkm}
 M.~Kawasaki, K.~Kohri and T.~Moroi,
 Phys.\ Rev.\ D {\bf 71} (2005) 083502
 [arXiv:astro-ph/0408426]; 
  K.~Jedamzik,
  Phys.\ Rev.\  D {\bf 74}, 103509 (2006)
  [arXiv:hep-ph/0604251],
 and references therein.
   
\bibitem{Gherghetta:1999sw}
  T.~Gherghetta, G.~F.~Giudice and J.~D.~Wells,
  Nucl.\ Phys.\  B {\bf 559}, 27 (1999)
  [arXiv:hep-ph/9904378].
\bibitem{hep-ph/9906527} 
  T.~Moroi and L.~Randall,
  Nucl.\ Phys.\ B\ {\bf 570}, 455  (2000)
  [hep-ph/9906527].
 
\bibitem{Ibe:2004tg}
  M.~Ibe, R.~Kitano, H.~Murayama and T.~Yanagida,
  Phys.\ Rev.\  D {\bf 70}, 075012 (2004)
  [arXiv:hep-ph/0403198];
  M.~Ibe, R.~Kitano and H.~Murayama,
  Phys.\ Rev.\  D {\bf 71}, 075003 (2005)
  [arXiv:hep-ph/0412200].
\bibitem{ArkaniHamed:2006mb} 
  N.~Arkani-Hamed, A.~Delgado and G.~F.~Giudice,
  Nucl.\ Phys.\ B {\bf 741}, 108 (2006)
  [hep-ph/0601041].
\bibitem{leptogenesis}
  M.~Fukugita and T.~Yanagida,
  Phys.~Lett.~{\bf B174} (1986) 45; 
  For  reviews,
  W.~Buchmuller, R.~D.~Peccei and T.~Yanagida,
  Ann.\ Rev.\ Nucl.\ Part.\ Sci.\  {\bf 55}, 311 (2005)
  [arXiv:hep-ph/0502169];  
  S.~Davidson, E.~Nardi and Y.~Nir,
  Phys.\ Rept.\ \ {\bf 466}, 105  (2008)
  [arXiv:0802.2962 [hep-ph]].
  
\bibitem{Feng:1999fu}
  J.~L.~Feng, T.~Moroi, L.~Randall, M.~Strassler and S.~f.~Su,
  Phys.\ Rev.\ Lett.\  {\bf 83}, 1731 (1999).
\bibitem{hep-ph/0610249} 
  J.~Hisano, S.~Matsumoto, M.~Nagai, O.~Saito and M.~Senami,
  Phys.\ Lett.\ B\ {\bf 646}, 34  (2007)
  [hep-ph/0610249].
\bibitem{arXiv:0706.4071} 
  M.~Cirelli, A.~Strumia and M.~Tamburini,
  Nucl.\ Phys.\ B\ {\bf 787}, 152  (2007)
  [arXiv:0706.4071 [hep-ph]].

  
\bibitem{Inoue:1991rk}
  K.~Inoue, M.~Kawasaki, M.~Yamaguchi and T.~Yanagida,
  Phys.\ Rev.\ D {\bf 45}, 328 (1992).
\bibitem{hep-th/0405159} 
  N.~Arkani-Hamed and S.~Dimopoulos,
  JHEP\ {\bf 0506}, 073  (2005)
  [hep-th/0405159].
\bibitem{hep-ph/0406088} 
  G.~F.~Giudice and A.~Romanino,
  Nucl.\ Phys.\ B\ {\bf 699}, 65  (2004)
  [Erratum-ibid.\ B\ {\bf 706}, 65  (2005)]
  [hep-ph/0406088].
\bibitem{hep-ph/0409232} 
  N.~Arkani-Hamed, S.~Dimopoulos, G.~F.~Giudice and A.~Romanino,
  Nucl.\ Phys.\ B\ {\bf 709}, 3  (2005)
  [hep-ph/0409232].
\bibitem{Izawa:2010ym} 
  K.~-I.~Izawa, T.~Kugo and T.~T.~Yanagida,
  Prog.\ Theor.\ Phys.\  {\bf 125}, 261 (2011)
  [arXiv:1008.4641 [hep-ph]].

\bibitem{hep-ph/0411041} 
  J.~D.~Wells,
  Phys.\ Rev.\ D\ {\bf 71}, 015013  (2005)
  [hep-ph/0411041].
\bibitem{arXiv:1111.4519} 
  L.~J.~Hall and Y.~Nomura,
  arXiv:1111.4519 [hep-ph].
\bibitem{TU-363} 
  Y.~Okada, M.~Yamaguchi and T.~Yanagida,
  Phys.\ Lett.\ B\ {\bf 262}, 54  (1991).


\bibitem{arXiv:0705.1496} 
  N.~Bernal, A.~Djouadi and P.~Slavich,
  JHEP\ {\bf 0707}, 016  (2007)
  [arXiv:0705.1496 [hep-ph]].
  
\bibitem{arXiv:1108.6077} 
  G.~F.~Giudice and A.~Strumia,
  arXiv:1108.6077 [hep-ph].
  \bibitem{arXiv:1107.5255} 
  M.~Lancaster [Tevatron Electroweak Working Group and for the CDF and D0 Collaborations],
  arXiv:1107.5255 [hep-ex].
\bibitem{arXiv:0908.1135} 
  S.~Bethke,
  Eur.\ Phys.\ J.\ C\ {\bf 64}, 689  (2009)
  [arXiv:0908.1135 [hep-ph]].
\bibitem{hep-ex/0306033} 
  R.~Barate {\it et al.} [LEP Working Group for Higgs boson searches and ALEPH and DELPHI and L3 and OPAL Collaborations],
  Phys.\ Lett.\ B\ {\bf 565}, 61  (2003)
  [hep-ex/0306033].

\bibitem{arXiv:1001.4538} 
  E.~Komatsu {\it et al.} [WMAP Collaboration],
  Astrophys.\ J.\ Suppl.\ \ {\bf 192}, 18  (2011)
  [arXiv:1001.4538 [astro-ph.CO]].
\bibitem{hep-ex/0203020} 
  A.~Heister {\it et al.} [ALEPH Collaboration],
  Phys.\ Lett.\ B\ {\bf 533}, 223  (2002)
  [hep-ex/0203020].
\bibitem{arXiv:0810.1892} 
  J.~Hisano, M.~Kawasaki, K.~Kohri and K.~Nakayama,
  Phys.\ Rev.\ D\ {\bf 79}, 063514  (2009)
  [Erratum-ibid.\ D\ {\bf 80}, 029907  (2009)]
  [arXiv:0810.1892 [hep-ph]].


\bibitem{hep-ph/0610277} 
  M.~Ibe, T.~Moroi and T.~T.~Yanagida,
  Phys.\ Lett.\ B\ {\bf 644}, 355  (2007)
  [hep-ph/0610277].

\bibitem{arXiv:0705.3086} 
  S.~Asai, T.~Moroi, K.~Nishihara and T.~T.~Yanagida,
  Phys.\ Lett.\ B\ {\bf 653}, 81  (2007)
  [arXiv:0705.3086 [hep-ph]].
\bibitem{arXiv:0802.3725} 
  S.~Asai, T.~Moroi and T.~T.~Yanagida,
  Phys.\ Lett.\ B\ {\bf 664}, 185  (2008)
  [arXiv:0802.3725 [hep-ph]].
\bibitem{Alves:2011ug} 
  D.~S.~M.~Alves, E.~Izaguirre and J.~G.~Wacker,
  arXiv:1108.3390 [hep-ph].
\bibitem{Fujii:2002kr} 
  M.~Fujii and K.~Hamaguchi,
  Phys.\ Rev.\ D {\bf 66}, 083501 (2002)
  [hep-ph/0205044];
  M.~Fujii and M.~Ibe,
  Phys.\ Rev.\ D\ {\bf 69}, 035006  (2004)
  [hep-ph/0308118].
  

\end{thebibliography}
\end{document}